\documentclass[3p,times]{elsarticle}
\usepackage{ecrc}
\usepackage{xcolor}
\usepackage{soul}
\usepackage{amssymb}
\usepackage[figuresright]{rotating}

\volume{00}
\firstpage{1}
\journalname{Journal of Molecular Liquids}
\runauth{V.M.~Pergamenshchik et al.}
\jid{JML}
\jnltitlelogo{JML}
\CopyrightLine{2011}{Published by Elsevier Ltd.}

\begin{document}

\begin{frontmatter}



\dochead{}

\title{Canonical partition function and distance dependent correlation
	functions of a quasi-one-dimensional system of hard disks}


\author[label1,label2]{V.M.~Pergamenshchik$^{*}$}, 
\author[label3,label4]{T. Bryk}  
\author[label3]{A. Trokhymchuk}
\address[label1]{Institute of Physics, National Academy of Sciences of Ukraine,
	prospekt Nauky, 46, Kyiv 03039, Ukraine, $^{*}$victorpergam@yahoo.com}
		
\address[label2]{Center for Theoretical Physics, Polish Academy of Sciences, Al. Lotnik\'{o}w 32/46, 02-668, Warsaw, Poland}
	
\address[label3]{Institute for Condensed Matter Physics, NAS of Ukraine, 
1 Svientsistsky Str, Lviv, 79011, Ukraine}

\address[label4] {Institute of Applied Mathematics and Fundamental
	Sciences,\\Lviv National Polytechnic University, UA-79013 Lviv,
	Ukraine }

\begin{abstract}
The canonical $\,NLT\,$ partition function of a quasi-one dimensional
(q1D) one-file system of equal hard disks [J. Chem Phys. \textbf{153}, 144111
(2020)] provides an analytical
 description of the thermodynamics and
ordering in this system (a pore) as a function of linear density $\,Nd/L\,$ 
where $\,d\,$ is the disk diameter.
We derive the analytical formulae for the distance dependence of the
translational pair distribution function and the distribution function of distances
between next neighbor disks, and then demonstrate their use by calculating
the translational order in the pore. In all cases, the order is found to be
of a short range and to exponentially decay with the disks' separation. The
correlation length presented for different pore widths and densities shows a
non-monotonic dependence with a maximum at $\,Nd/L = 1\,$ and tends to
the 1D value for a vanishing pore width. The results indicate a special role
of this density when the pore length $\,L\,$ is equal exactly to $N$ disk
diameters. Comparison between the theoretical results for an infinite system and the results of a molecular dynamics simulation for a finite system with periodic boundary conditions is presented and discussed. 
\end{abstract}


\end{frontmatter}



\section{Introduction}

\label{Int}

The statistical description of many-particle systems must deal with many,
even infinite number of degrees of freedom and as many integrals. As this
limit can be studied only theoretically, the analytical results and
particularly exact ones are of great importance. To solve a statistical
mechanical problem implies to reduce the problem of calculation of its
partition function (PF) and pair correlation functions to a finite number of
dimensions, finite number of integrals and other mathematical actions. This
is most often a task impossible and we try to learn the physics of many-body
system and develop the appropriate mathematical tools by studying simplified
models. In particular, a strong simplification can be achieved by
considering geometries with reduced dimensionality and, in particular,
one-dimensional (1D). A great number of 1D models considered in the last
century and summarized in the book \cite{Lieb} has proved to be very
usefully related to the physics in two and three dimensions. In the theory
of liquids, modeling molecules as hard spheres, the distinguished example of
the 1D physics is the exact solution for the PF of a 1D gas of {\ hard core
molecules}, now known as Tonks' gas \cite{Tonks}.

The 1D Tonks gas is much simpler than any 2D system, nevertheless Tonks'
solution has become the analytical platform for further expansion into the
world of 2D hard disk (HD) systems via moving to certain
quasi-one-dimensional (q1D) models. The simplest q1D HD system is such that
each disk can touch no more than one next neighbor from both sides (the
so-called single-file system); the width of such q1D pore must be below ($%
\sqrt{3}/2+1)d$ where $d$ is the HD diameter. The analytical theory of HDs
in q1D pore was first considered by Wojciechowski et al \cite{Wojc} for a
system periodically replicated in the transverse direction. Later Kofke and
Post \cite{Kofke} proposed an approach that enables one to consider HDs in a
q1D pore in the thermodynamic limit using the well-known transfer matrix
method introduced in statistical physics by Kramers and Wannier \cite%
{Kramers}. This theory has become the main tool in studying a q1D
single-file HD system \cite{Varga,Varga2013,Godfrey2015,Robinson,Hu,Comment}
that allowed one to address also q1D systems of non-circular particles
(e.g., \cite{squares}). The virial expansion for disks in the q1D geometry
has been addressed by Mon \cite{Mon}. Thanks to the analytical methods
nowadays HDs in the q1D geometry have been intensively used for calculations
of thermodynamic as a model glass former to study glass transitions and HDs'
dynamics \cite{Robinson,Yamchi,Hicks,AiT}. The approximate analytical theory
for HDs was also developed for one- and two-dimensional random pore geometry
by the scaled particle method \cite{Holovko}. The new interest has been
brought about by the studies of actual physical ultracold systems such as
Bose-Einstein condensates created in practically 1D or q1D electromagnetic
traps \cite{BEC}. Although mathematically quantum and classical gases are
very different, the classical 1D and q1D models can provide some technical
and even physical insight.

The transfer matrix method is essentially related to the pressure-based $%
\,NPT\,$ ensemble which does not directly predict pressure as a function of
system's width $\,D\,$ and length $\,L\,$\ as the Gibbs free energy is
parameterized by the pressure $\,P$. Another peculiarity is that the
transfer matrix approach essentially employs the periodic boundary
conditions along the pore which allows one to reduce the PF to the trace of
the transfer matrix \cite{Kofke}. Recently one of us derived analytically
the canonical $\,NLT\,$ PF of a q1D HD single-file system (from now on q1D
implies also single-file system) both for a finite number of disks $\,N\,$
and in the thermodynamic limit \cite{JCP2020}. Although the derivation of
the PF in \cite{JCP2020} has used certain approximation which is addressed
in more detail in the next section, the obtained \ results provide
convenient analytical tools. In particular, the PF of a $NLT$ ensemble is
not related to the periodic boundary conditions and finding the
thermodynamic properties of a q1D HD system for given $\,L\,$ and $\,D\,$ is
reduced to solving single transcendental equation which can be easily done
numerically. The PF, pressure along and across the pore, distribution of the
contact distances between neighboring HDs along the pore, and distribution
of HD centers across the pore for given linear density $\rho =Nd/L$ are
found analytically. In this paper we derive and employ another fundamental
thermodynamic quantity, the pair distribution function (PDF).

The transfer matrix method directly gives the leading correlation length
that describes the correlations between the disks' transverse coordinates $%
\,y_{i}$ and $y_{i+n}\,$\ as a function of the difference $n$ between their
order numbers \cite{Varga,Varga2013,Godfrey2015,Robinson,Hu,Comment}. At the
same time, the most important longitudinal ordering is related to PDFs that
are functions of {the actual distance} $\,R\,$ between disks along the
system. The analytical formula for {such PDF }$g_{1D}(R)$ is known only for
a 1D gas of noninteracting (Tonks' gas) \cite{Frenkel,Yukhnovski,Santos} and
interacting $\ $\cite{Santos} hard core molecules . In a q1D system, finding
the PDF $\,g(R)\,$for large $R$ by the transfer matrix method directly from
its definition is admittedly a formidable problem \cite{Varga2013}. The
large $R$ behavior of the PDF is possible to get by means of the following
nontrivial numerical procedures: either by inverse Fourier transform of the
structure factor obtained from the joint solution of two integral equations 
\cite{Robinson}, or by first planting the system's configuration from the
transfer matrix eigenstates and then averaging over these planted
configurations \cite{Comment}. The main goal of this paper is to develop an
alternative, analytical approach to the PDFs of a q1D HD system based on the 
$\,NLT\,$ ensemble PF, which is not related to periodic boundary conditions
along the pore, and demonstrate its implementation.

From the analytical canonical PF of a q1D HD system~\cite{JCP2020}, we
derive a formula for the translational PDF $\,g(R)\,$ which requires
computing a few integrals and can be straightforwardly implemented
numerically. We also derive the PDF $\,$for the distance between next
neighbors. Both PDFs are presented for an infinite system, but the canonical
PF allows one to obtain the formulae for finite systems, too. Usually, the
PDF for a 1D gas is derived by resorting to Laplace's transform related to
the NPT ensemble \cite{Yukhnovski,Santos}, but in earlier works Frenkel \cite%
{Frenkel} and Nagamija \cite{Nagamiya} used a more direct technique related
to a converting infinite products into exponentials. We also use the last
technique and derive the PDF for a q1D HD system directly from the canonical
PF. The method is first demonstrated by application to the PDF of a 1D Tonks
gas and the derived formulae then used to calculate the translational PDF $%
\,g(R)\,$ and its correlation length for the range of the q1D pore widths
and wide range of linear densities of a q1D HD system. In all cases, the
correlations are found to exponentially decay with the disks' separation.
The correlation length presented for the total range of the q1D pore widths
and different densities shows a non-monotonic density dependence with a
maximum at the density $\,Nd/L=1\,$ and, for vanishing pore width, tends to
the 1D value of a Tonks gas. The theoretical PDFs $\,g(R)\,$ and $%
\,g_{1}(R)\,$ are compared with the results of molecular dynamic (MD)
simulations presented for q1D systems of $\,N=400\,$ and 2000 HDs.\emph{\ }%
It is found that the theoretical and computer simulations results for $%
g_{1}(R)$ nearly coincide for high and low densities. At the same time, at
the intermediate densities in the vicinity of $\,\rho \sim 1\,$, to coincide
with the MD results, the theoretical $g_{1}$ should be obtained for a
slightly higher density.\emph{\ }Analysing MD simulations data for PDF $%
\,g(R)\,$, we came to a tentative conclusion that this difference can be
attributed to the approximation used in \cite{JCP2020} and, possibly, to a
pressure difference between a finite system with periodic boundary condition
considered in computer simulations and an infinite system considered in the
theory.

The paper is structured as follows. The canonical PF and the methods of its
calculations are introduced in Sec.~2, and then, in Sec.~3, the formulas for
PDFs $\,g(R)\,$ and $\,g_{1}(R)\,$ are derived. In Sec.~4\thinspace , these
formulae are used to study the theoretical PDFs, the results are compared
with the MD data and discussed in detail. The final Sec.5 is a brief
conclusion.

\section{The canonical partition function of a q1D HD system}

\label{Sec2}

Consider a pore of length $\,L,$ confined between two parallel hard walls
separated by the width $\,D,\,$ and filled with $\,N\,$ of HDs of diameter $%
d=1$. All lengths will be measured in units of HD diameters. The reduced
width $\,\Delta =(D-d)/d\,$, that gives the actual pore width attainable to
HD centers, in the single-file q1D case ranges from $0$ in the 1D case to
the maximum $\,\sqrt{3}/2\approx 0.866\,$. The $\,i$-th disk has two
coordinates, $\,x_{i}\,$ along and $\,y_{i}\,$ across the pore; $\,y\,$
varies in the range $\,-\Delta /2\leq y\leq \Delta /2\,$; the pore volume is 
$\,LD\,$. The transverse center-to-center distance between two neighbors, $%
\,\delta y_{i}=y_{i+1}-y_{i}\,$, determines the contact distance $\,\sigma
\, $ between them along the pore\thinspace : 
\begin{eqnarray}
\sigma (\delta y_{i}) &=&\min \left\vert
x_{i+1}(y_{i+1})-x_{i}(y_{i})\right\vert ,  \nonumber \\
\sigma (\delta y_{i}) &=&\sqrt{d^{2}-\delta y_{i}^{2}},  \label{sigma} \\
\sigma _{m}(\Delta ) &=&\sqrt{d^{2}-\Delta ^{2}}\leq \sigma \leq d\,. 
\nonumber
\end{eqnarray}

The minimum possible contact distance, $\,\sigma _{m}\,$, depends on the
pore width $\,\Delta \,$ and obtains for $\,\delta y_{i}=\pm \Delta \,$ when
the two disks are in contact with the opposite walls. Thus, each set of
coordinates $\,\{y\}=y_{1},y_{2},\ldots ,y_{N}\,$ determines the
correspondent densely packed state of the total length $\,L^{\prime
}\{y\}=\sum_{i=1}^{N-1}\sigma (\delta y_{i})\,$, which we call condensate 
\cite{JCP2020}. The minimum condensate length (the distance between centers
of the first and $N$ th disk) is $\,(N-1)\sigma _{m}\,$, the maximum length
can be as large as $\,(N-1)d\,$, but it cannot exceed $\,L-d\,$, i.e., $%
\,(N-1)\sigma _{m}<L^{\prime }\leq L_{\max }^{\prime }\,$ where $\,L_{\max
}^{\prime }=\min [(N-1)d,L-d]\,$.

The PF of this q1D HD system\ is given by the integral \cite{JCP2020}: 
\begin{equation}
Z=\Delta \int\nolimits_{\displaystyle-\infty }^{\displaystyle\infty }\frac{%
d\alpha }{N!}\int_{\displaystyle(N-1)\sigma _{m}}^{\displaystyle %
L_{m}^{\prime }}dL^{\prime }e^{\displaystyle i\alpha L^{\prime
}}(L-1-L^{\prime })^{N}\left( \int_{\displaystyle\sigma _{m}}^{\displaystyle%
1}\frac{d\sigma }{\sqrt{1-\sigma ^{2}}}\sigma e^{\displaystyle-i\alpha
\sigma }\right) ^{N-1},  \label{Z}
\end{equation}%
where $\,i\,$ is an imaginary unite. To derive the above PF, in \cite%
{JCP2020}, the integration over transverse coordinates $y$\ was changed to
that over $\sigma (\delta y)$. The integration over different $\delta y_{i}$%
\ is not independent and so is the integration over $\sigma (\delta y_{i}),$%
\ but in \cite{JCP2020} this integration was performed for each $\sigma
(\delta y_{i})$\ from $\sigma _{m}$\ to $1$ independently. It turns out that
the PF obtained under the above approximation coincides with that obtained
in \cite{Wojc} for a system periodic in the transverse direction. However,
in \cite{JCP2020} such periodic condition in $y$\ was not imposed. The above
approximation is supposed to be valid in the limit of large $N$\ as in this
limit the main contribution to the PF has to come from the average contact
distance $\overline{\sigma }$ which does lie within the range from $\sigma
_{m}$\ to 1. This approximation has allowed one to solve the problem
analytically for the system in a box and avoid the periodic boundary
condition in $x$.

It is convenient to rewrite this PF in the exponential form\thinspace : 
\begin{equation}
Z=\frac{\Delta }{N!}\int_{\displaystyle(N-1)\sigma _{m}}^{\displaystyle %
L_{m}^{\prime }}dL^{\prime }\int d\alpha e^{\displaystyle S}\,,  \label{Z3}
\end{equation}%
where 
\begin{equation}
S=i\alpha L^{\prime }+N\ln (L-1-L^{\prime })+(N-1)\ln \left( \int_{%
\displaystyle\sigma _{m}}^{\displaystyle1}\frac{d\sigma }{\sqrt{1-\sigma ^{2}%
}}\sigma e^{\displaystyle-i\alpha \sigma }\right) \,.  \label{S}
\end{equation}

Equations~(\ref{Z3}) and (\ref{S}) give the PF in the general case of a q1D
HD system for large $\,N\,$ and $\,L\,$. \ 

The integrand of $\,Z\,$ is a regular function of $\,\alpha \,$ so that the $%
\,\alpha $-integration contour, in particular, its central part that gives
the principal contribution to the integral, can be shifted while the ends
remain along the real axis. In the thermodynamic limit $\,N\rightarrow
\infty \,$, $\,L\rightarrow \infty \,,Nd/L=\rho =const\,$, we can compute
the PF (\ref{Z3}) by the steepest descent method. In the limit $N\rightarrow
\infty $ the integral (\ref{Z3}) is exactly determined by the saddle point
which, for given $N\,$, $L\,$ and $\,\sigma _{m}\,$, is the stationary point
of the function $\,S(i\alpha ,L^{\prime })\,$, Eq.~(\ref{S}). It is
convenient to introduce real $\,a=i\alpha \,$ since $\,\alpha \,$ at the
saddle point lies on the imaginary axis and the integration contour has to
be properly deformed. The two equations $\,\partial S/\partial a=\partial
S/\partial L^{\prime }=0\,$ that determine the saddle point, can be reduced
to the single equation for $\,a=a_{N}\,$ which reads\thinspace : 
\begin{equation}
\frac{L}{N}-\frac{1}{a_{N}}=\frac{I^{\prime }(a_{N})}{I(a_{N})}\,,  \label{E}
\end{equation}%
where 
\begin{eqnarray}
I(a_{N}) &=&\int_{\sigma _{m}}^{1}\frac{d\sigma }{\sqrt{1-\sigma ^{2}}}%
\sigma \exp (-a_{N}\sigma )\,,  \label{I} \\
&&  \nonumber \\
I^{\prime }(a_{N}) &=&\int_{\sigma _{m}}^{1}\frac{d\sigma }{\sqrt{1-\sigma
^{2}}}\sigma ^{2}\exp (-a_{N}\sigma )\,.  \label{I1}
\end{eqnarray}%
The solution $\,a_{N}\,$ of Eq.~(\ref{E}), which gives the total
longitudinal pressure $\,P_{L}=k_{B}Ta_{N}/D$ \cite{JCP2020} and
longitudinal force $\,k_{B}Ta_{N}\,$ \cite{AiT} where$\,k_{B}T$ is
Boltzmann's constant times temperature, depends on the per disk pore length $%
\,L/N\,$ and, via $\sigma _{m}\,$, on the pore width $\,D\,$, and fully
determines the free energy. The free energy per disk,$\,F/N$, which
therefore is the function of the pore length $\,L\,$, pore width $\,D\,$ and
the temperature $\,T\,$, is $\,F(L,D,T)/N=-TS(a_{N})/N=-Ts_{N}\,$ where $%
\,s_{N}\,$ is system's per disk entropy\thinspace : 
\begin{equation}
s_{N}=a_{N}\sigma _{N}+\ln \left( L-N\sigma _{N}\right) +\frac{N-1}{N}\ln
I(a_{N})\,,  \label{S1}
\end{equation}%
where $\sigma _{N}$ is the average value of the contact distance $\,\sigma
\, $ in the condensate [i.e., average of $\,L^{\prime }/(N-1)\,$] \cite%
{JCP2020}: 
\begin{equation}
\sigma _{N}=\frac{L}{N}-\frac{1}{a_{N}}\,.  \label{sigN}
\end{equation}%
Finally, for $\,N\rightarrow \infty \,$, the PF can be cast in the two
equivalent forms\thinspace : 
\begin{eqnarray}
Z_{\displaystyle\infty } &=&\frac{\varsigma _{N}\Delta }{N!}\exp (Ns_{N})
\label{ZZZ} \\
&=&\frac{\varsigma _{N}\Delta }{N!}(L-N\sigma _{N})^{N}I(a_{N})^{N-1}\exp
(Na_{N}\sigma _{N})\,,  \nonumber
\end{eqnarray}%
where $\,\varsigma _{N}\,$ is the prefacfor $\,\sim 1/\sqrt{N}\,$ originated
from the Gaussian integration along the steepest descent contour whose exact
form is of no need. In the 1D case, all $\sigma $'s are equal to $\,d\,$ and
this expression goes over into the Tonks PF $\,Z_{1D}\,$ up to the factor $%
\,\Delta ^{N}\,$ which in this case represents the independent transverse
degrees of freedom: $\,Z_{\infty }\rightarrow \Delta ^{N}Z_{1D}\,$ where 
\begin{equation}
Z_{1D}=\frac{1}{N!}\left( L-Nd\right) ^{N}\theta \left( L-Nd\right) \,
\label{Z1D}
\end{equation}%
and $\,\theta \left( x\right) \,$ is the step function equal to 1 for $%
\,x>0\,$ and $\,0\,$ otherwise. Now consider the general case of a finite
system. In what follows, the number of HDs and the total length of a finite
system are denoted as $\,n\,$ and $\,L_{n}\,$, respectively (instead of $%
\,N\,$ and $\,L\,$). The integral (\ref{Z}) can be transformed to the one
along the real axis $\alpha \,$like that\thinspace : 
\begin{eqnarray}
Z_{\displaystyle n,L_{n}} &=&\frac{\Delta }{n!}\int_{\displaystyle{%
(n-1)\sigma _{m}}}^{\displaystyle{L_{m}}}dL^{\prime }\left(
L_{n}-1-L^{\prime }\right) ^{\displaystyle{n}}  \nonumber \\
&&  \label{Z1} \\
&&\times \int_{\displaystyle-\infty }^{\displaystyle\infty }\frac{%
\displaystyle{d\alpha }}{\displaystyle{2\pi }}\left\vert I(i\alpha
)\right\vert ^{\displaystyle{n-1}}\cos \left[ L^{\prime }\alpha
+(n-1)\varphi _{\alpha }\right] \,,  \nonumber
\end{eqnarray}%
where $\,L_{m}=\min (n-1,L_{n}-1)\,$ and 
\begin{eqnarray}
\varphi _{\alpha } &=&\arg I(i\alpha ),  \nonumber \\
&&  \label{bII} \\
I(i\alpha ) &=&\int_{\sigma _{m}}^{1}\frac{d\sigma }{\sqrt{1-\sigma ^{2}}}%
\sigma e^{\displaystyle{-i\alpha \sigma }}\,.  \nonumber
\end{eqnarray}

Although the Gaussian approximation at the saddle point cannot give an exact
result for a system with finite number of disks, choosing the $\,\alpha \,$
integration contour passing through the saddle point provides the best
convergence of the integrals (which has been confirmed numerically). Hence
to compute the PF we shift the central part of the $\,\alpha \,$ integration
contour downward and integrate over the real variable $t$ along the line $%
\,\alpha =-ia_{n}+t\,$ that crosses the imaginary axis at $\,\alpha
=-ia_{n}\,$. The best choice for the shift $\,a_{n}\,$ is the root of the
following modified equation (\ref{E}): 
\begin{equation}
\frac{L_{n}}{n}-\frac{n-1}{na_{n}}=\frac{I^{\prime }(a_{n})}{I(a_{n})}\,,
\label{En}
\end{equation}%
whose rhs is defined in Eqs.~(\ref{I}) and (\ref{I1}). Then the PF $%
\,Z_{n,L_{n}}\,$ can be transformed like that: 
\begin{eqnarray}
Z_{n,L_{n}} &=&\frac{\Delta }{n!}\int_{{(n-1)\sigma _{m}}}^{{L_{m}}%
}dL^{\prime }e^{\displaystyle{a_{n}L^{\prime }}}\left( L_{n}-1-L^{\prime
}\right) ^{n}  \nonumber \\
&&  \label{ZnR} \\
&&\times \int_{-\infty }^{\infty }\frac{dt}{2\pi }\left(
I_{s}{}^{2}+I_{c}^{2}\right) {}^{(n-1)/2}\cos [L^{\prime }t+(n-1)\varphi ]\,.
\nonumber
\end{eqnarray}%
where 
\begin{eqnarray}
I_{s}(t) &=&-\int_{\sigma _{m}}^{1}\frac{d\sigma }{\sqrt{1-\sigma ^{2}}}%
\sigma e^{\displaystyle{-a_{n}\sigma }}\sin (t\sigma )\,,  \nonumber \\
&&  \label{IcIs} \\
I_{c}(t) &=&\int_{\sigma _{m}}^{1}\frac{d\sigma }{\sqrt{1-\sigma ^{2}}}%
\sigma e^{\displaystyle{\ -a_{n}x}}\cos (t\sigma )\,,  \nonumber
\end{eqnarray}%
\begin{equation}
\varphi (t)=\arg \left( I_{c}+iI_{s}\right) =\left\{ 
\begin{array}{c}
\arctan \displaystyle{\frac{I_{s}}{I_{c}}}\,\,,I_{c}>0\,, \\ 
\pi +\arctan \displaystyle{\frac{I_{s}}{I_{c}}}\,\,,I_{c}<0,I_{s}>0\,, \\ 
-\pi +\arctan \displaystyle{\frac{I_{s}}{I_{c}}}\,\,,I_{c}<0,I_{s}<0\,.%
\end{array}%
\right. \,.  \label{arg2}
\end{equation}%
The density $\,n/L_{\displaystyle n}\,$ and the reduced pore width $\,\Delta
\,$, which enter the integrals above via $\,\sigma _{m}\,$, fully determine
the partition function $\,Z_{\displaystyle n,L_{\displaystyle n}}\,$ through
Eqs.~(\ref{En}) - (\ref{arg2})\thinspace .

\section{Derivation of the PDF from the canonical partition function}

\label{Sec3}

The PDF as a function of separation $\,R\,$ is the probability to find
particle a distance $\,R\,$ from another particle whose coordinate $%
\,x_{0}\, $ is fixed, say at $\,x_{0}=0\,$. Here we derive $\,g(R)\,$ for a
q1D HD systems directly from the PF {$\,Z_{N}\,$} of the canonical $\,NLT\,$
ensemble.

The q1D PF $\,Z_{N}\{x_{i},y_{i}\}\,$, Eqs.~(\ref{Z3}) and (\ref{S}), is a
functional of the particles' longitudinal $\,x\,$ coordinates and transverse 
$\,y\,$ coordinates. In the particular case of a q1D system, the general
formula for the PDF $\,g(R)\,$ equivalent to its definition is obtained from
the canonical PF for the $\,N\,$ particle system by fixing the $\,x\,$
coordinate of $\,n$-th disk at $\,x_{n}=x\,$ and then summing over all
possible $\,n\,$ (the range of $\,n\,$ will be clarified later on)\,: 
\begin{equation}
g(R)=\frac{1}{\rho }\sum_{n=1}\frac{Z_{N}%
\{x_{0}=0,y_{0},x_{1},y_{1},...,x_{n}=R,y_{n},...,x_{N},y_{N}\}}{%
Z_{N}\{x_{0}=0,y_{0},x_{1},y_{1},...,x_{n},y_{n},...x_{N},y_{N}\}}\,.
\label{P(R)}
\end{equation}
Note that $\,y_{0}\,$ and $\,y_{n}\,$ are not fixed so that the particles $%
\,0\,$ and $\,n\,$ can move in the transverse direction. The PF in the
nominator splits into a product of two PFs, $\,Z_{n}\,$ for $\,n\,$ disks
(of which $\,n-1\,$ are free to move) in the space $\,0<x_{k}<R\,$, and $%
\,Z_{N-n, L-R}\,$ for $\,N-n\,$ moving disks in the space $%
\,R<x_{k}<L-R-d/2\,$, Fig.~1\thinspace : 
\begin{equation}
g(R)=\frac{1}{\rho }\sum_{n=1}\frac{Z_{n,R}Z_{N-n,L-R}}{Z_{N,L}}\,.
\label{P1}
\end{equation}
Figure~1 demonstrates that the numbers of free disks, contact distances $%
\sigma \,$, and {the disk-vertical wall contact distance } $\,d/2\,$ have to
be adjusted in each PF individually. As a result, the form of Eq.~(\ref{E})
that determines $\,a_{n}\,$, is also slightly modified.

\begin{figure}[tbp]
\centering
\includegraphics[width=0.75\textwidth]{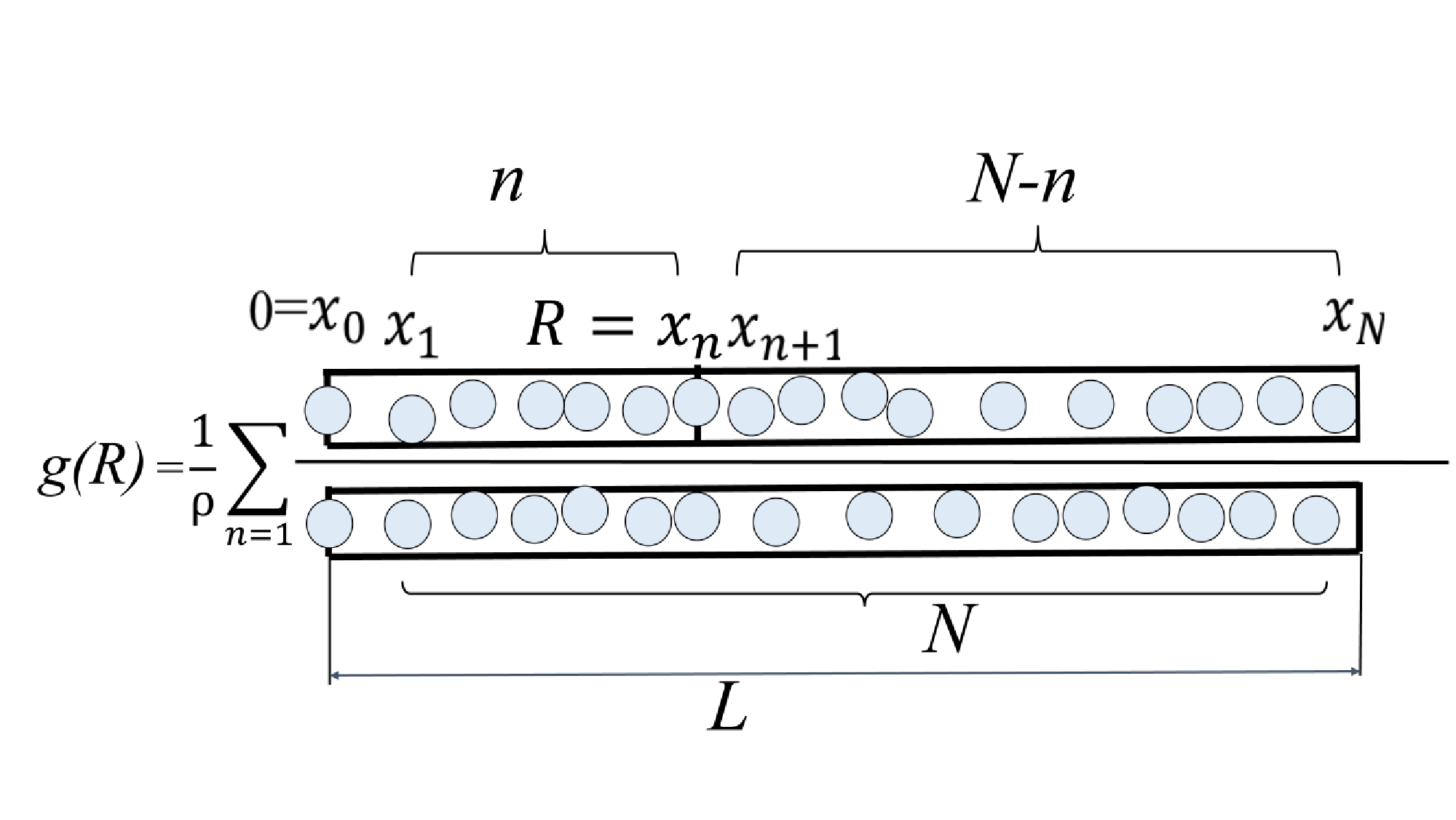}
\caption{Definition of the pair distribution function $\,g(R)\,$ and the
three PFs: $\,Z_{n, R}\,$ is for $\,n-1\,$ free moving disks in the space $%
\,R\,$ in which there are $\,n\,$ neighbors of disk $\,0\,$; $\,Z_{N-n,
L-R}\,$ is for $\,N-n\,$ free disks in the space $\,L-R\,$; and $\,Z_{N}\,$
is for $\,N\,$ free disks in the space $\,L\,$.}
\label{gR}
\end{figure}

Consider first $\,Z_{n,R}\,$. We assume that $\,R>1\,$; the case $\,R<1\,$,
possible only for $\,n=1\,$, will be considered separately. In the system of
size $\,R>1\,$, there are $\,n-1\,$ freely moving HDs, {\ }$\int
dy_{0}=\Delta $, $n$ contact distances $\sigma $, {no vertical walls and
thus no contact distances } $\,d/2\,$. Hence $(R-d-L^{\prime })^{n}$ in PF (%
\ref{ZnR}) has to be replaced by $\,(R-L^{\prime })^{n-1}$ and $I(i\alpha
)^{n-1} $ by $I(i\alpha )^{n}\,$. Then the PF $\,Z_{n,R}\,$ takes the form 
\begin{eqnarray}
Z_{n,R} &=&\frac{\Delta }{(n-1)!}\int_{\displaystyle n\sigma _{m}}^{%
\displaystyle L_{n,m}}dL^{\prime }e^{\displaystyle{a_{n}L^{\prime }}}\left(
R-L^{\prime }\right) ^{\displaystyle n-1}  \label{ZZn} \\
&&  \nonumber \\
&&\times \int_{\displaystyle-\infty }^{\displaystyle\infty }\frac{dt}{2\pi }%
(I_{s}{}^{2}+I_{c}{}^{2})^{\displaystyle n/2}\cos [L^{\prime }t+n\varphi
(t)]\,,  \nonumber
\end{eqnarray}
where $\,L_{n,m}=\min (n,R)\,$. The angle $\,\varphi (t)\,$ is defined in
Eq.~(\ref{arg2}), $\,a_{n}\,$ is the root of equation (\ref{En}), and the
relation between $\sigma _{n}$ and $a_{n}$ is just the properly modified
Eq.~(\ref{sigN})\,, 
\begin{equation}
\sigma _{n}=\frac{R}{n}-\frac{n-1}{na_{n}}\,.  \label{b}
\end{equation}

Next consider $\,Z_{N-n,L-R}\,$ in Eq.~(\ref{P1}), the PF for $\,N-n\,$ HDs
in the range $\,R<x<L\,$. Here all disks are free to move, there are $%
\,N-n\, $ contact distances $\,\sigma\,$, and the {single vertical wall }at
the pore end. Then the PF $\,Z_{N-n,L-R}\,$ can be presented in the form 
\begin{eqnarray}
Z_{\displaystyle N-n,L-R} &=&\frac{1}{(N-n)!}\int_{\displaystyle\sigma
_{m}}^{\displaystyle L_{N-n,m}}dle^{ \displaystyle{L^{\prime }a_{N-n}}%
}(L-R-1/2-L^{\prime })^{\displaystyle N-n}  \nonumber \\
&&  \label{ZN-n} \\
&&\times \int_{\displaystyle-\infty }^{\displaystyle\infty }\frac{dt}{2\pi }%
(I_{s}{}^{2}+I_{c}{}^{2})^{\displaystyle(N-n)/2}\cos [L^{\prime
}t+(N-n)\varphi (t)]\,,  \nonumber
\end{eqnarray}
where $\,L_{\displaystyle N-n, m}=\min (N-n,L-R-1/2)$ and $a_{\displaystyle %
N-n}$ is the root of the following modified Eq.~(\ref{E})\,: 
\begin{equation}
\frac{L-R-1/2}{N-n}-\frac{1}{a_{N-n}}=\frac{I^{\prime }(a_{\displaystyle %
N-n})}{I(a_{\displaystyle N-n})}\,.  \label{EN-n}
\end{equation}

At last, consider $\,Z_{N, L}\,$ in Eq.~(\ref{P1}) that is the PF for $\,N\,$
HDs in the range $0<x<L\,$. Here all disks are free to move, $\int
dy_{0}=\Delta \,$, there are $N$ contact distances $\sigma $ and the{\
single vertical wall} at the pore end. Then the PF $\,Z_{N,L}\,$ can be
presented in the form 
\begin{eqnarray}
Z_{N,L} &=&\frac{\Delta }{N!}\int_{\displaystyle N\sigma _{m}}^{%
\displaystyle L_{N,m}}dL^{\prime }e^{ \displaystyle{a_{N}L^{\prime }}%
}(L-1/2-L^{\prime })^{N}  \nonumber \\
&&  \label{ZNL} \\
&&\times \int_{\displaystyle-\infty }^{\displaystyle\infty }\frac{dt}{2\pi }%
(I_{s}{}^{2}+I_{c}{}^{2})^{\displaystyle N/2}\cos [L^{\prime }t+N\varphi
(t)]\,,  \nonumber
\end{eqnarray}
where $\,L_{\displaystyle N,m}=\min (N,L-1/2)\,$ and $\,a_{N}\,$ is the root
of the equation (\ref{E}). Making use of the PFs (\ref{ZZn}), (\ref{ZN-n}),
and (\ref{ZNL}) in the general formula (\ref{P1}) gives the PDF of a q1D HD
system for finite $\,N\,$ and $\,L\,$.

The general result for $\,g(R)\,$ can be further simplified in the
thermodynamic limit. This case, usually considered the most important one,
is presented in detail in the next section. To illustrate our method of
deriving the PDF directly from the canonical $\,NLT\,$ PF we first derive
the PDF for 1D Tonks' gas.

\subsection{PDF of a q1D HD system in the thermodynamic limit}

\subsubsection{PDF of an infinitely long 1D HD system (Tonks' gas)}

The PDF $\,g(R)\,$ for a 1D HD is given by the general formula (\ref{P1}) in
which the three PFs are obtained from the Tonks' PF, Eq.~(\ref{Z1D}%
)\thinspace : 
\begin{equation}
g_{1D}(R)=\frac{1}{\rho }\sum_{n=1}\frac{N!|R-n|^{n-1}[L-R-n]^{N-n}}{%
(n-1)!(N-n)!(L-n)^{N}}\,\theta (R-n)\,.  \label{g1D}
\end{equation}%
In the limit $N\rightarrow \infty \,$, neglecting $\,O(n/N)\,$, one also has 
$\,(N-n)!\cong N!/N^{n}\,$ and 
\begin{eqnarray}
(L-R-n)^{N-n} &=&(L-N)^{N-n}\left( 1-\frac{R-n}{N(l_{N}-1)}\right) ^{N-n} 
\nonumber \\
&=&(L-N)^{N-n}\left[ \exp \left( -\frac{R-n}{l_{N}-1}\right) +O\left( \frac{n%
}{N}\right) \right]  \label{e1} \\
&\rightarrow &(L-N)^{N-n}\exp \left( -\frac{R-n}{l_{N}-1}\right) \,, 
\nonumber
\end{eqnarray}
where $\,l_{N}=L/N=1/\rho \,$. Making use of these results in Eq.~(\ref{g1D}%
) and introducing the step function, $\,\theta (R-n)=1\,$ for $\,R\geq n\,$
and $\,\theta (R-n)=0\,$ otherwise, one finally obtains 
\begin{equation}
g_{1D}(R)=\frac{1}{\rho }\sum_{n=1}\frac{|R-n|^{n-1}\exp \left( \displaystyle%
{\ \ -\frac{R-n}{l_{N}-1}}\right) }{(n-1)!(l_{N}-1)^{n}}\,\theta (R-n)\,,
\label{gg1D}
\end{equation}
which is the well-known PDF of 1D Tonks' gas \cite{Yukhnovski,Santos}.

\subsubsection{PDF of an infinitely long q1D HD system.}

In an infinitely long q1D HD system, the above thermodynamic limit result,
Eqs.~(\ref{Z3}) and (\ref{S}), is applicable both for $\,Z_{\displaystyle N,
L}\,$ and $\,Z_{\displaystyle N-n, L-R}\,$ as the number of particles $%
\,N-n\,$ and volume $\,L-R\,$ are infinite, but the PF $\,Z_{\displaystyle %
n,R}\,$ for the finite $\,n\,$ disk system has to be found directly from the
general formula (\ref{ZnR}) [or from the original form (\ref{Z1}) without
the contour shift]. Adjusting Eqs.~(\ref{S1})-(\ref{ZZZ}) to the above PFs
of interest, one has\,: 
\begin{eqnarray}
Z_{N-n,L-R} &=&\frac{\varsigma _{N-n}}{(N-n)!}[L-R-(N-n)\sigma
_{N-n}]^{N-n}\exp [(N-n)\widetilde{s}_{N-n}]\,,  \nonumber \\
&&  \label{ZZ} \\
Z_{N,L} &=&\frac{\varsigma _{N}\Delta }{N!}(L-N\sigma _{N})^{N}\exp (N%
\widetilde{s}_{N})\,.  \nonumber
\end{eqnarray}
Here $\widetilde{s}_{\displaystyle N-n}=a_{\displaystyle N-n}\sigma_{%
\displaystyle N-n}+\ln I(a_{N-n})\,$ and $\,\widetilde{s}_{\displaystyle %
N}=a_{\displaystyle N}\sigma_{\displaystyle N}+\ln I(a_{\displaystyle N})$,
where the pair $\sigma_{\displaystyle N}, a_{\displaystyle N}\,$ is
determined by $\,l_{N}=L/N=1/\rho \,$ from the Eqs.~(\ref{E}) and (\ref{sigN}%
) and the pair $\sigma_{\displaystyle N-n}, a_{\displaystyle N-n}\,$ by $%
\,l_{\displaystyle N-n}=(L-R)/(N-n)\,$ from similar equations 
\begin{eqnarray}
l_{N-n}-\frac{1}{a_{N-n}} &=&\frac{I^{\prime }(a_{N-n})}{I(a_{N-n})}\,, 
\nonumber \\
&&  \label{sigN-n} \\
\sigma _{N-n} &=&l_{N-n}-\frac{1}{a_{N-n}}\,.  \nonumber
\end{eqnarray}%
Substituting these expressions in the general formula (\ref{P1}) for $%
\,g(R)\,$ and taking into account that in the thermodynamic limit the
preexponential factors $\varsigma_{\displaystyle N}$ and $\varsigma_{%
\displaystyle N-n}$ are equal, we get\,: 
\begin{eqnarray}
g(R) &=&\frac{1}{\rho }\sum_{n=1}\frac{Z_{n,R}N![L-R-(N-n)\sigma
_{N-n}]^{N-n}}{(N-n)!(L-N\sigma _{N})^{N}}  \nonumber \\
&&  \label{P2} \\
&&\times \exp [N(\widetilde{s}_{N-n}-\widetilde{s}_{N})-n\widetilde{s}%
_{N-n}]\,.  \nonumber
\end{eqnarray}%
Now we find $\widetilde{s}_{N-n}$ by expanding about $\widetilde{s}_{N}$ and
using the smallness of $\ n/N.$ First, up to $O(n/R),$ one has $N(\widetilde{
s}_{N-n}-\widetilde{s}_{N})\cong N\left( \partial \widetilde{s}_{N}/\partial
l_{N}\right) (l_{N-n}-l_{N})\,$, where 
\begin{eqnarray}
l_{N-n}-l_{N} &=&\frac{L-R}{N-n}-\frac{L}{N}  \nonumber \\
&&  \label{del I} \\
&=&\frac{R-l_{N}n}{N}\left[ 1+O(n/L)\right] \,.  \nonumber
\end{eqnarray}%
The $\,l_{N}\,$ derivative obtains regarding (\ref{E}) and (\ref{sigN}%
)\thinspace : 
\begin{equation}
\frac{\partial \widetilde{s}_{N}}{\partial l_{N}}=\frac{1}{a_{N}}\frac{
\partial a_{N}}{\partial l_{N}}+a_{N}=a_{N}\frac{\partial \sigma _{N}}{
\partial l_{N}}\,.  \label{dsN}
\end{equation}%
Then one expands $\widetilde{s}_{N-n}$ about $\widetilde{s}_{N}$ regarding (%
\ref{del I})\thinspace : 
\begin{eqnarray}
n\widetilde{s}_{N-n} &\cong &n\widetilde{s}_{N}+n\frac{\partial \widetilde{s}
_{N}}{\partial l_{N}}(l_{N-n}-l_{N})  \label{Ns1} \\
&=&n\widetilde{s}_{N}+O(n/N)\,,  \nonumber
\end{eqnarray}%
to finally obtain 
\begin{equation}
N(\widetilde{s}_{N-n}-\widetilde{s}_{N})\cong -a_{N}\frac{\partial \sigma
_{N}}{\partial l_{N}}(R-nl_{N})\,.  \label{Nds}
\end{equation}%
Next we show that the $N-n$ th power of the ratio in (\ref{P2}) gives rise
to an exponential: 
\begin{eqnarray}
&&\left[ \frac{L-R-(N-n)\sigma _{N-n}}{L-N\sigma _{N}}\right] ^{N-n} 
\nonumber \\
&=&\left( \frac{l_{N}-\sigma _{N-n}}{l_{N}-\sigma _{N}}\right) ^{N-n}\left(
1-\frac{R-n\sigma _{N-n}}{L-N\sigma _{N-n}}\right) ^{N-n}  \label{exp} \\
&\cong &\left( \frac{l_{N}-\sigma _{N-n}}{l_{N}-\sigma _{N}}\right)
^{N-n}\exp \left( -\frac{R-n\sigma _{N}}{l_{N}-\sigma _{N}}\right) . 
\nonumber
\end{eqnarray}%
In turn, the first factor in the last line can also be reduced to an
exponential whose exponent cancels out the one in Eq.~(\ref{Nds})\thinspace
: 
\begin{eqnarray}
\left( \frac{l_{N}-\sigma _{N-n}}{l_{N}-\sigma _{N}}\right) ^{N-n} &=&\left(
1+\frac{\sigma _{N}-\sigma _{N-n}}{l_{N}-\sigma _{N}}\right) ^{N-n}\cong 
\nonumber \\
&&  \label{exp2} \\
\left[ 1+\frac{a_{N}}{N}\frac{\partial \sigma _{N}}{\partial l_{N}}\right]
^{N-n} &\cong &\exp \left[ a_{N}\frac{\partial \sigma _{N}}{\partial l_{N}}%
(R-nl_{N})\right] \,.  \nonumber
\end{eqnarray}%
Making use of the results (\ref{Nds})-(\ref{exp2}) in formula (\ref{P2}),
after some straightforward algebra and convenient rescaling, we obtain the
PDF in the final form: 
\begin{equation}
g(R)=\frac{1}{\rho }\sum_{\displaystyle n=1}^{\displaystyle n_{\max }}\,%
\frac{|R-n\sigma_{\displaystyle n}|^{\displaystyle n-1}\exp \left\{ %
\displaystyle -\frac{{R-n\sigma _{N}}}{{l_{N}-\sigma_{N} }} + n\left[ a_{%
\displaystyle n}\sigma_{\displaystyle n}-a_{N}\sigma_{N}+\ln \frac{{I(a_{%
\displaystyle n}) }}{\displaystyle{I(a_{N})}}\right] \right\} }{%
(n-1)!(l_{N}-\sigma_{N})^{\displaystyle n}}\,J_{\displaystyle n}(R)\,.
\label{PDF}
\end{equation}
Here $\,J_{n}(R)\,$ is the following integral: 
\begin{eqnarray}
J_{n}(R) &=&n\int_{\displaystyle{\sigma _{m}}}^{\displaystyle{l_{m}}}dle^{%
\displaystyle na_{n}(l-\sigma _{n})}\left( \frac{R/n-l}{|R/n-\sigma _{n}|}%
\right) ^{ \displaystyle{n-1}}  \nonumber \\
&&  \label{Jn} \\
&&\times \int_{-\infty }^{\infty }\frac{dt}{2\pi }\left[ \frac{
I_{c}(t)^{2}+I_{s}(t)^{2}}{I(a_{n})^{2}}\right] ^{\displaystyle{n/2}}\cos
[n(lt+\varphi )]\,,  \nonumber
\end{eqnarray}
where $\,I_{c}(t),I_{s}(t),\varphi (t)\,$ and $\,a_{n},\sigma _{n}\,$ are
given in Eqs.~(\ref{IcIs}), (\ref{arg2}) and Eqs.~(\ref{En}), (\ref{b}),
respectively. Deriving Eqs.~(\ref{PDF}) and (\ref{Jn}), we changed from the
variable $L^{\prime }$\ to $l=L^{\prime }/n$\ so that the upper $l$\
integration limit is now $l_{m}=\min (1,R/n).$\ To avoid dealing with
extremely small quantities and extremely fast oscillations, we made the
following convenient rescaling: we divided $R/n-l$ by $|R/n-\sigma _{n}|$\
and, to compensate, introduced the factor $|R-n\sigma _{n}|^{n-1};$\
similarly, the factor $\,\exp [na_{n}\sigma _{n}+n\ln I(a_{n})]\,$
compensates for the denominator $I(a_{n})^{{n}}\,$ and $\,\exp
[-na_{n}\sigma _{n}]\,$ in the integrand.

The maximum 
$n_{\max }$ in summation of Eq.~({\ref{PDF}}) is the maximum number of disks
at close contact which can be put in the space between the particle fixed at 
$\,x=0\,$ and the point $\,x=R\,$: 
\begin{equation}
n_{\max }(R)=\frac{R-\mathrm{mod}(R,\sigma_{m})}{\sigma_{m}}\,.
\label{n max}
\end{equation}
Note that the expression for $\,g(R)\,$ appears to be considerably simpler
if no contour shift and rescaling have been applied: 
\begin{eqnarray}
g(R) &=&\frac{1}{\rho }\sum_{{\displaystyle n=1}}^{{\displaystyle n_{\max}}}%
\frac{n\exp \left( \displaystyle{-\frac{R-n\sigma_{N}}{l_{N}-\sigma_{N}}}%
\right) }{ (n-1)!(l_{N}-\sigma _{N})^{\displaystyle n}}  \nonumber \\
&&  \label{gsimpl} \\
&&\times \int_{\displaystyle\sigma_{m}}^{\displaystyle l_{m}}dl\left(
R/n-l\right) ^{n-1}\int_{\displaystyle -\infty }^{\displaystyle\infty }\frac{%
d\alpha }{2\pi }\left\vert I(i\alpha )\right\vert ^{\displaystyle n/2}\cos
[n(l\alpha +\varphi_{\alpha })]\,,  \nonumber
\end{eqnarray}
where $\,I(i\alpha)\,$ and $\,\varphi_{\alpha}\,$ are defined in (\ref{bII}%
). But the formulae (\ref{PDF}) and (\ref{Jn}) actually provide a much
better convergence and much simpler numericals.

\subsubsection{The 1D limit}

It is important to see how the results obtained for a q1D HD system behave
approaching a 1D HD system, i.e., in the limit $\,D\rightarrow 0\,$ when $%
\,\Delta \rightarrow 0\,$, and $\,\sigma_{m}\,, \sigma_{n}\,, \sigma_{N}
\rightarrow 1\,$. To this end, we first estimate the $\,x\,$ integrals in
this limit\,: 
\begin{eqnarray}
I(a) &=&e^{\displaystyle -a}\Delta +O(\Delta ^{2})\,,  \nonumber \\
I_{c} &=&e^{\displaystyle -a}\Delta \cos t+O(\Delta ^{2})\,,  \label{I0} \\
I_{s} &=&e^{\displaystyle -a}\Delta \sin t+O(\Delta ^{2})\,.  \nonumber
\end{eqnarray}
As a result, 
\begin{eqnarray}
\varphi &\rightarrow &-t\,,  \nonumber \\
&&  \nonumber \\
\frac{I_{\displaystyle c}^{\displaystyle 2}+I_{\displaystyle s}^{%
\displaystyle 2}}{I(a_{\displaystyle n})^{\displaystyle 2}} &\rightarrow
&1\,,  \nonumber \\
&&  \nonumber \\
\displaystyle\int_{\displaystyle -\infty }^{\displaystyle\infty }\frac{dt}{%
2\pi }\left[ \frac{I_{\displaystyle c}^{\displaystyle 2}+I_{\displaystyle %
s}^{\displaystyle 2}}{ I(a_{\displaystyle n})^{\displaystyle 2}}\right] ^{%
\displaystyle n/2}\cos [n(lt+\varphi )] &\rightarrow &\delta (l-1)\,,
\label{delta(l-1)} \\
&&  \nonumber \\
J_{\displaystyle n} &\rightarrow &1\,.  \nonumber
\end{eqnarray}
We see that in the 1D limit, the $\,g(R)\,$, Eq.~(\ref{PDF}), goes over into
the Tonks $\,g_{1D}(R)\,$, Eq.~(\ref{gg1D}).

\subsubsection{Probability to find next neighbor at a distance $R$}

The term with $\,n=1\,$ in the PDF $\,g(R)\,$ is proportional to the
probability $\,g_{1}(R)=Z_{1,R}Z_{N-1,L-R}/\rho Z_{N,L}\,$ to have next
neighbour disk $\,1\,$ of disk $\,0\,$ at a distance $\,R\,$ including $%
\,R<1\,$. Here we derive this quantity for an infinitely long q1D HD system.
The case $\,n=1\,$ is particular because the small distance between neighbor
disks sets certain restriction on the integration over their transverse
coordinates $\,y\,$ which depends on their distributions. Now we have to
consider the two neighbor disks, $\,0\,$ and 1, within the large system. The
result is similar to that obtained in \cite{JCP2020} in deriving the $\,y\,$
distribution across the pore. This distribution has the form $\,\propto
\varphi (y)^{2}\,$ where $\,\varphi \,$ is the following integral\thinspace
: 
\begin{equation}
\varphi (y)=\int_{-\Delta /2}^{\Delta /2}dy^{\prime }e^{\displaystyle-\alpha
_{N}\sigma (y-y^{\prime })}.  \label{fi}
\end{equation}%
Here $\sigma (y-y^{\prime })$ is defined in eq.(\ref{sigma}), and, compared
with formulae of~\cite{JCP2020}, the integration variable $\sigma $ is
changed to $y$. This derivation shows that to place the two disks into a
large system, it is sufficient to consider correlations between disk $0$ and
one disk on the left of disk $0$, call it disk $-1,$ and that between disk 1
and one disk, call it disk 2, on the right of disk $1.$ Then, rather than
disks $0$ and 1 we consider disks $-1,0,1,$ and 2 which results in the
following extensions:%
\begin{eqnarray*}
\int_{-\Delta /2}^{\Delta /2}dy_{0} &\rightarrow &\int_{-\Delta /2}^{\Delta
/2}dy_{-1}\int_{-\Delta /2}^{\Delta /2}dy_{0}e^{\displaystyle-\alpha
_{N}\sigma (y_{0}-y_{-1})}=\int_{-\Delta /2}^{\Delta /2}dy_{0}\varphi
(y_{0}), \\
&& \\
&& \\
\int_{-\Delta /2}^{\Delta /2}dy_{1} &\rightarrow &\int_{-\Delta /2}^{\Delta
/2}dy_{1}\int_{-\Delta /2}^{\Delta /2}dy_{2}e^{\displaystyle-\alpha
_{N}\sigma (y_{2}-y_{1})}=\int_{-\Delta /2}^{\Delta /2}dy_{1}\varphi (y_{1}).
\end{eqnarray*}%
Regarding the equalities $\sigma _{N-1}=\sigma _{N+1}$ and $\widetilde{s}%
_{N-1}=\widetilde{s}_{N+1}$ valid up to $O(1/N)$ and retaining only the $%
\,R\,$ dependent terms $z(R)$, one has: 
\begin{equation}
z(R)=\int_{-\Delta /2}^{\Delta /2}dy_{0}\int_{-\Delta /2}^{\Delta
/2}dy_{1}\varphi (y_{0})\varphi (y_{1})\theta \left[
R^{2}+(y_{0}-y_{1})^{2}-1\right] \exp \left( -a_{N}R\right) ,  \label{z}
\end{equation}%
where the $\theta $ function eliminates states in which the cores of disks $0
$ and $1$ overlap and we used that $1/(l_{N}-\sigma _{N})=a_{N}$.
Normalizing on unity, one finally obtains: 
\begin{equation}
g_{1}(R)=\frac{z(R)}{\displaystyle{\int_{\sigma _{m}}^{\infty }dRz(R)}}\,.
\label{gg1}
\end{equation}%
In relation with the approximation used to derive the PF (\ref{Z3}) and
described in Sec.2, we stress that the distribution $\varphi (y)^{2}$\ with $%
\varphi $\ given in ( \ref{fi}) is very different from that in the system
periodic in the $y$\ direction \cite{Kofke}, and the approximation
influences this distribution only through the value of $a_{N}.$

\section{Results and discussion}

\label{Sec4}

Figures 2 and 3 present the PDF $\,g_{1}(R)$ for next neighbor disks
obtained from Eq.~(\ref{gg1}) for a set of linear densities $\,\rho =N/L\,$
and two reduced pore widths $\Delta $. The sharp peak at $\,R=1\,$ is
present at all densities including very high, but in this case its height is
incomparable with {second peak} centered at the average interdisk spacing $%
\,l_{N}=1/\rho \,$. {\ The second peak} appears and strengthens as density
becomes higher and higher. The concentration of spacings $\,R\,$ at the
average distance indicates {a high order} along the pore. For densities $%
\,\rho \,$ near the close packing, this also implies {a high overall zigzag
order} since $\,R\cong l_{N}\,$ approaches the minimum separation $\,\sigma
_{m}\,$ for which disks stay very close to the walls. In contrast, the fact
that there is a high peak at $\,R=1\,$ which is particularly pronounced for
the density $\,\rho =1\,$ with $\,l_{N}\,=1$ shows that the ordering at this
density is not necessarily related with a zigzag type order. We shall give
this issue a more consideration later on as the peculiarity of separation $%
\,R=1\,$ and density $\,\rho =1\,$ will get additional indications.\emph{\ }%
Right now we would like only to explain the very reason for the cusp at $%
\,R=1\,$ whose presence at PDFs $\,g_{1}(R)\,$ and $\,g(x)\,$ has been well
known \cite{Varga2013,Godfrey2015,JCP2020,WE,Comment}. To this end, the $%
\theta $ function in the formula for $z(R)\,$, Eq.~(\ref{z}), is replaced by
the explicit dependence of the integration limits on $R,$ i.e.,%
\begin{equation}
z(R)=\left\{ 
\begin{array}{c}
\displaystyle{\int_{-\Delta /2}^{\Delta /2-\sqrt{1-R^{2}}}dy_{0}\int_{y_{0}+%
\sqrt{1-R^{2}}}^{\Delta /2}dy_{1}\varphi (y_{0})\varphi (y_{1})},\,\,\,R\leq
1\,, \\ 
\\ 
\displaystyle{\int_{-\Delta /2}^{\Delta /2}dy_{0}\int_{-\Delta /2}^{\Delta
/2}dy_{1}\varphi (y_{0})\varphi (y_{1})},\,\,\,R>1\,.%
\end{array}%
\right.  \label{zz}
\end{equation}%
This formula shows that the increase of the disk transverse free path in $%
\,y\,$ with distance $\,R\,$ abruptly stops at its maximum constant value $%
\,\Delta \,$ at $\,R=1$.

\begin{figure}[tph]
\centering
\includegraphics[width=0.475\textwidth]{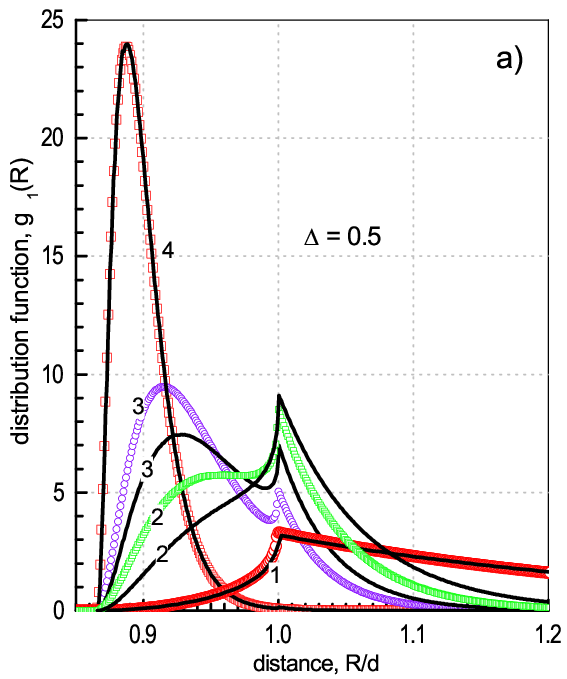}
\par
\includegraphics[width=0.65\textwidth]{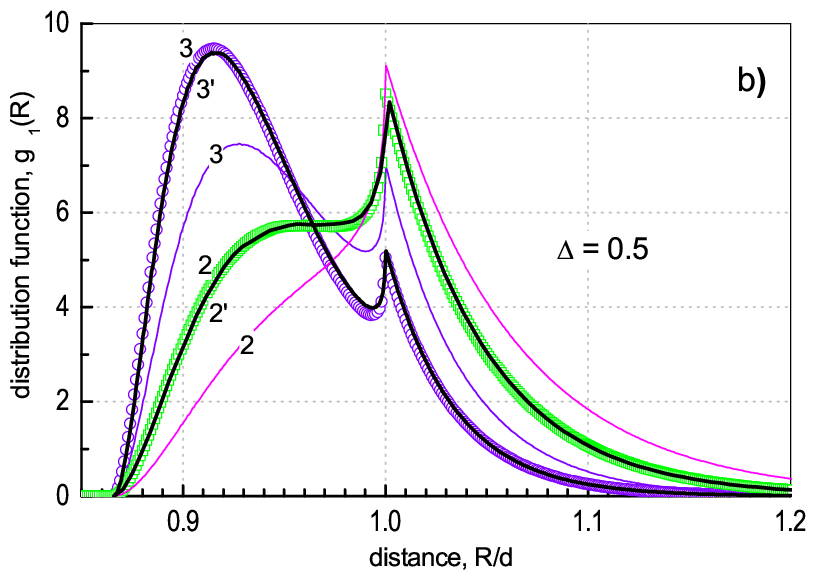}
\caption{Part a): Theoretical results (solid lines) and MD simulation data
(symbols) for pair distribution function $\,g_{1}(R)\,$ for pore width $%
\,\Delta =0.5\,$ and four densities: 1 -- $\,\protect\rho =0.8\,$;\, 2 --
1.01;\, 3 -- 1.053;\, 4 -- 1.111\,. Part b): Theoretical results for shifted
densities $\,\protect\rho =1.034\,$ and $\,\protect\rho =1.065\,$ (thick
solid lines) which are practically indistinguishable of the MD simulation
data (symbols) for densities $\,\protect\rho=1.01\,$ (right peak) and $\,%
\protect\rho =1.053\,$ (leftt peak), respectively. For comparison, the thin
solid lines 2 and 3 [the same as in part a)] show theoretical curves for
actual (non shifted) densities $\,\protect\rho =1.01\,$ and 1.053\,,
respectively\,.}
\label{Fig2}
\end{figure}

\begin{figure}[tph]
\centering
\includegraphics[width=0.475\textwidth]{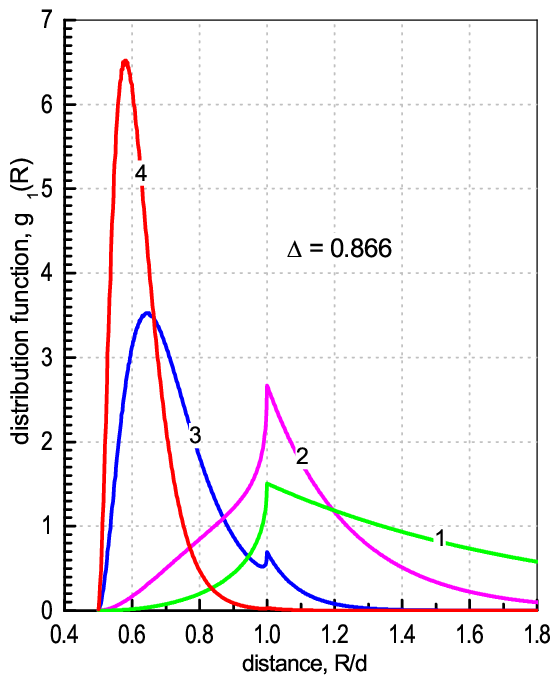}
\caption{Theoretical pair distribution function $\,g_{1}(R)\,$ for pore
width $\,\Delta =0.866\,$ and four different densities: 1 -- $\,\protect\rho%
=0.6\,$; 2 -- 1;\thinspace\ 3 -- 1.4 and 4 -- 1.6\,.}
\label{Fig3}
\end{figure}

In Fig.~2a, the theoretical $\,g_{1}(R)\,$ is superimposed on the MD
simulation data~\cite{WE} for the same $\,\Delta =0.5\,$. It is seen that
the theoretical and MD simulation results for high $\,\rho =1.111\,$ and low 
$\,\rho =0.8\,$ practically coincide whereas for the intermediate densities $%
\,\rho=1.053\,$ and 1.01 they look very different. Actually, however, a
perfect fit can be achieved by a small increase of these theoretical
densities respectively to $1.065$ and $1.032$, Fig.~2b. This mismatch is
addressed after presenting the case of pore width $\,\Delta =0.866\,$ in
Fig.~3. This figure shows our theoretical $\,g_{1}(R)\,$ for the densities
for which in \cite{Varga2013} the PDF normalized on the density, $g/\rho ,$
was obtained by a Monte Carlo simulation for short distances $R<1.5$. While $%
\,g_{1}(R)\,$ contains the contribution of a single next neighbor, $\,g(R)\,$
in \cite{Varga2013} includes the contributions of both next and next-next
neighbor. Nevertheless, the peaks for $\rho =1.6$ and $1.4$ are concentrated
at distances $R<1$ where the contribution of the next-next neighbor is
indirect and negligible so that our $g_{1}/\rho $ and the $g/\rho $ can be
compared. After dividing by the correspondent $\rho ,$ these curves in Fig.3
become in a good agreement with their counterparts from \cite{Varga2013}.
The results for $\,\rho =1\,$ and 0.6, however, cannot be compared as the
role of the next-next neighbor for these curves in \cite{Varga2013} is
essential.

The first idea is that the reason for the aforementioned mismatch lies in
the approximation described in Sec.2 which was used in the derivation of the
PF in \cite{JCP2020}. This can be checked by developing the theory which
does not use both above approximation and periodic boundary condition in the 
$x$\ direction. At the same time, to address the mismatch between our
theoretical and MD results at the intermediate densities for $\,\Delta =0.5\,
$, it is also important to resort to Fig.~4 which presents our theoretical $%
g(R)$ and the MD results for $\,\rho =1\,$. Figure~4a shows that the general
trend is that the theoretical peaks obtained for an infinitely long system
are slightly wider and lower than those obtained by the MD simulations for a
system with periodic boundary conditions. The theoretical correlation length 
$\,1/0.124\approx 8.1\,$ is slightly shorter than that of the fit to MD data 
$\,1/0.1015\approx 9.85\,$, Fig.~4b. This last figure, however, also
demonstrates a visually appreciable difference between the theoretical and
MD $\,g(R)\,$: the last does not vanish for large $\,R\,$ and remains at a
level on the order of 0.01 for both system sizes $\,N=400\,$ and $N=2000\,$.
The residual correlations persist for all $R\gtrsim 50,$ are highly
fluctuating, showing no tendency to decreasing and are even higher for the
larger system. This points to the possibility that the slower correlation
decay of MD data is connected to the system finite size,\textbf{\ }i.e., the
effect which was already addressed in \cite{Comment}. Another reason can be
the periodic boundary conditions employed in MD simulations: imposing a
correlation at the distance equal to the system length $\,L\,$ can also
enforce the longitudinal correlation value. The relation between theoretical
and computer simulation data was already addressed in \cite{AiT} where we
compared the data for the transverse disk distributions. It was found that
the former always predict less disks at the walls, i.e., at $\,R\sim \sigma
_{m}\,$, and slightly more disks at a distance $\,R\sim 1\,$ than the latter$%
.$ The reason is that the space for windowlike defects with $\,R\sim 1\,$ is
related to the system size $\,L\,$: it diminishes sharper with the linear
density $\,\rho \,$ for shorter $\,L\,$ and for sufficiently high $\,\rho \,$
in a finite size system is not available at all. At the same time, the
probability of a window $\,R\sim 1\,$ in the zigzag arrangement in an
infinite q1D system is nonzero for any $\,\rho \,$ below close packing~\cite%
{JCP2020}. For this reason it can be expected that the peaks of $\,g_{1}(R)\,
$ and $\,g(R)\,$ in an infinite theoretical system are slightly stretched
toward higher $\,R\,$, hence are wider and slightly lower than those
obtained in computer simulations of a finite system with periodic boundary
conditions, which is the case of Fig.4. In terms of correlation decay, this
means that an infinite system has a shorter correlation length than that in
a finite system with periodic boundary conditions. Clearly, in terms of
pressure and density it implies that the pressure sensitively depends on the
ordering details: in a finite system pressure is slightly higher than in an
infinite system and the mismatch between the two PDFs can be eliminated by
shifting the density of an infinite system to a slightly higher value, which
was demonstrated in Fig.2b. Another question is why this shift is mostly
needed at the intermediate densities, i.e., those between the dense packing
and gas values. Before addressing this we first consider the results
presented in Figs.~5 and 6.

\begin{figure}[tph]
\includegraphics[width=0.475\textwidth]{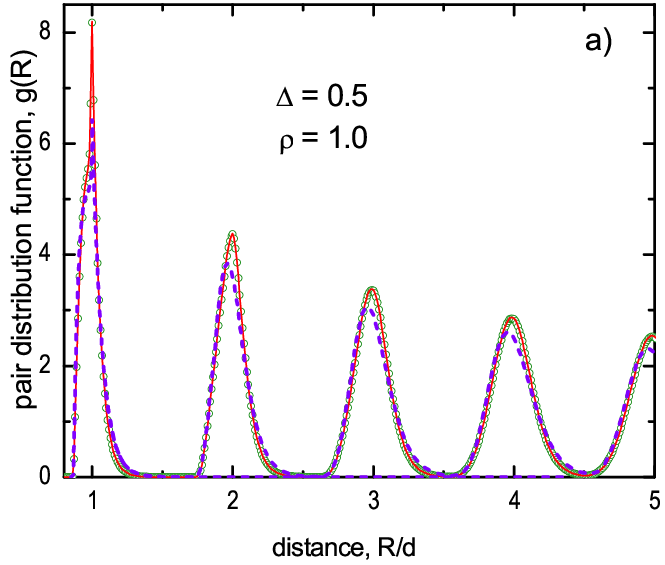} %
\includegraphics[width=0.5\textwidth]{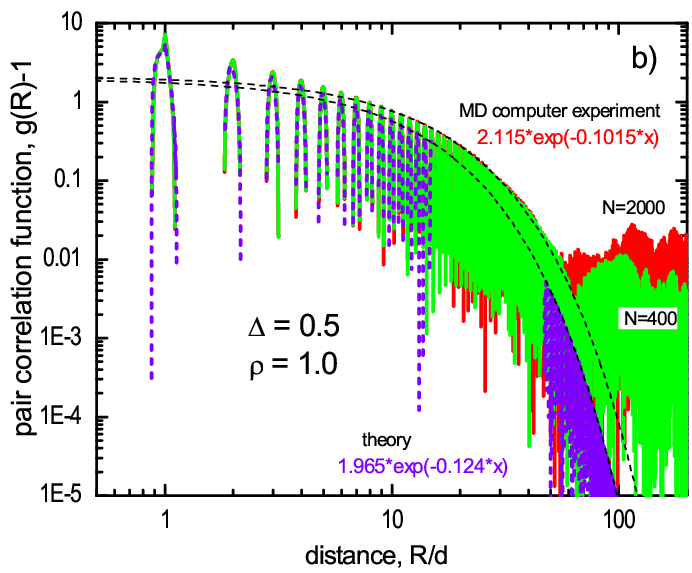}
\caption{Theoretical PDF $\,g(R)\,$ (the dashed line) superimposed on MD
simulation data (the solid line and symbols) for the case of pore width $%
\,\Delta =0.5\,$ and density $\,\protect\rho =1\,$. The MD data are shown
for two distinct sizes of the simulated system, i.e., $\,N=400\,$ (the green
color) and $\,N=2000\,$ (the red color). Part a) for distances $\,R<5\,$ and
part b) for distances $\,R<200\,$. In part b) the theoretical curve is
cropped in the range $\,15<R/d<50\,$ for better visualization of the
simulation data. }
\end{figure}

\begin{figure}[tph]
\centering
\includegraphics[width=0.75\textwidth]{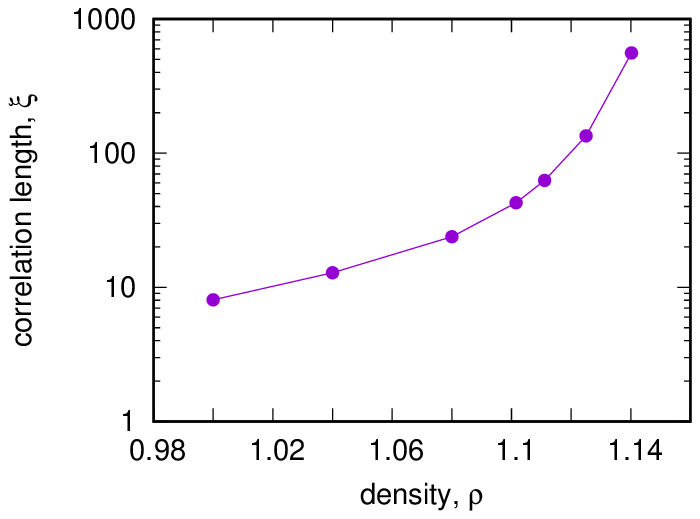}
\caption{Dependence of the correlation length $\,\protect\xi\,$ on density $%
\,\protect\rho\,$ for the case of pore width $\,\Delta =0.5\,$.}
\end{figure}

The longitudinal pair correlations as function of the disks' number
difference, $\,g_{2}(|n_{2}-n_{1}|)\,$, has been investigated in detail by
the transfer matrix method \cite{Varga,Godfrey2015,Robinson,Hu,Comment}. At
the same time, the PDF $\,g(R)\,$ as function of the disk separation $\,R\,$
for given density cannot be directly obtained by this method. Formula (\ref%
{PDF}) considerably simplifies its calculation and enables one to get its
systematic understanding by means of the direct calculation. The density $%
\,\rho \,$ determines $\,a_{N}\,$ (i.e., the pressure) via simple
transcendental Eq.~(\ref{E}) in which $\,\Delta \,$ enters via minimum
contact distance $\,\sigma _{m}\,$, Eq.~(\ref{sigma}), and $\,\sigma _{N}\,$
is given by Eq.~(\ref{sigN}). We obtained the PDF $\,g(R)\,$, Eq.~(\ref{PDF}%
), by performing the integration in Eq.~(\ref{Jn}) numerically, Figs. 5 and
6 \cite{ArXiv}. Contrary to our suggestion based on computer simulations
data~\cite{WE} and in line with the results of the transfer matrix method 
\cite{Comment}, our findings on the longitudinal correlations in the
thermodynamic limit show an exponential decay for all pore widths and
densities. The correlation length is a monotonically increasing function of
density, Fig. 5. To combine both width and density effects, we fixed the
ratio $\,\left( \rho /\rho _{\max }\right) \,$ of the actual density $\,\rho
\,$ to the maximum density $\,\rho _{\max }(\Delta )\,$ for a given pore
width $\,\Delta \,$, and then found the correlation lengths for different $%
\,\Delta \,$ in the total range of the single-file widths, $\,0\leq \Delta
\leq \sqrt{3}/2\approx 0.866\,$, Fig.~6. For a given $\,\Delta \,$ the
maximum density is $\,\rho _{\max }(\Delta )=1/\sigma _{m}(\Delta )=1/\sqrt{%
1-\Delta ^{2}}\,$. It follows that as $\,\Delta \,$ runs from $\,0\,$ to
0.866, the actual density $\rho =$ $(\rho /\rho _{\max })/\sqrt{1-\Delta ^{2}%
}$ monotonically increases from $0$ to $1.\,\allowbreak 154\,7(\rho /\rho
_{\max })$. In particular, for the same $\Delta \,$, the actual $\rho \,$ is
higher for higher $\rho /\rho _{\max }\,$. The results for $(\rho /\rho
_{\max })=0.866\,$, 0.9539 and 0.9875 are presented in Fig.~6\thinspace .

First, it is seen that, for the same $\,\Delta\,$, the correlation length is
larger for a higher density. Second, as the width approaches zero, the
correlation length tends to the value obtained for the 1D Tonks gas from $%
\,g_{1D}(R)\,$, Eq.~(\ref{g1D}). Third, the width and density monotonically
grow along the curves in Fig.\~{6}. It is seen however that the correlation
length does not monotonically {\ increase} as both the width and density do:
there is a maximum at each of the three curves. But the most interesting
observation is that all these maxima occur at the density $\,\rho =1\,$ when
a pore length interval equal to the disk diameter $\,d\,$ is on average
occupied by one disk. This is another peculiarity of these density and
disks' separation indicated above. 
\begin{figure}[tph]
\includegraphics[width=0.75\textwidth]{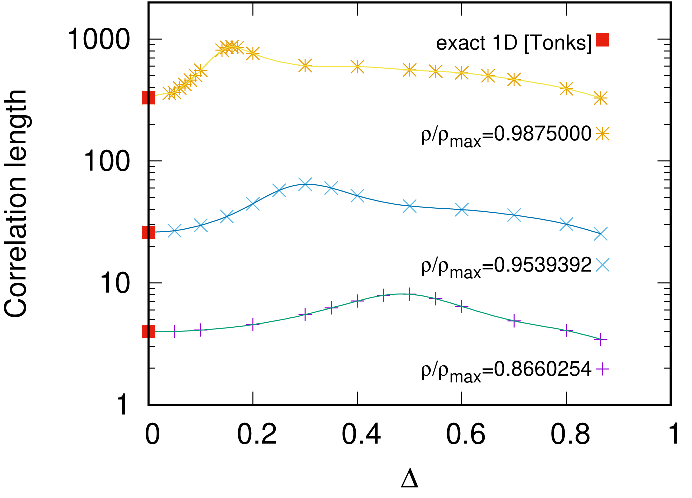}
\caption{Numerically obtained correlation length (symbols) for the
correlation function $\,g(R)-1\,$ as a function of width $\,\Delta\,$ for
the fixed ratio $\,(\protect\rho / \protect\rho_{\max })\,$ [for given pore
width $\,\Delta\,$, the maximum density $\,\protect\rho_{\max }=1/\protect%
\sigma_{m}(\Delta)\,$]. The three curves correspond to the three different
ratios $\,(\protect\rho / \protect\rho_{\max })\,$ indicated in the figure.
All maxima on different curves appear at density $\,\protect\rho =1\,$.}
\label{Fig6}
\end{figure}

Thus, Figs. 2, 3, and 6 show that the PDFs $g_{1}(R)$ and $g(R)$ have
peculiarities at the density $\rho =1$ in the form of certain peaks or
maxima. Moreover, for $\Delta =0.5,$ density $\rho =1$ is in the
intermediate range between high and low densities where we found a high
sensitivity of the pressure to density. Consider this effect which can be
related to the mismatch between the correlations in an infinite and finite
system with periodic boundary condition. It has been suggested that the peak
at the distribution of next neighbors at $R=1$, Figs.2, 3, is related to the
tendency of the system to produce windowlike defects to increase the entropy
as such a defect enables disk's travel across the pore \cite%
{Saika,Robinson,JCP2020,WE}. However, our finding that the correlation
length has a maximum at $\,\rho =1\,$ for any pore width, Fig.6, is
unexpected and cannot be explained by this idea alone. At higher densities,
the peak at $\,R=1\,$ is diminishing and the peak at another distinguished,
namely average distance $\,l_{N}=L/N<1\,$ is raising and eventually
dominates the one at $\,R=1\,$. As the peak of $\,g_{1}(R)\,$ at $\,R=L/N\,$
is definitely related to the longitudinal component of the zigzag order, it
is natural to connect the peak at $\,\rho =1\,$, at least partially, to the
nascent longitudinal ordering, too. In the light of this idea, the maxima of
the correlation length become reminiscent of the correlation length increase
at a phase transition. Of course, there is no transition at $\,\rho =1\,$,
but a kind of pretransitional effect seems to show up. Interestingly, in
recent paper on the same q1D HD system \cite{Santos1}, the authors found,
also for all widths $\Delta $'s, well developed compressibility peaks at $%
\rho \approx 1$\ showing that at this density the system is softer even than
at lower densities, which is in line with the above idea. We may then
speculate that at $\,\rho =1\,$, this effect is somehow related to the
increase of the correlation length and to enforced correlation sensitivity
at densities in the vicinity of $\rho =1.$ Approaching thermodynamic
equilibrium, a system tends to find more space to increase its entropy. For
densities near close packing, it has no much choice: the interdisk space is
very limited, correlating many such spacings along the system requires
extremely fine adjustments so that the correlations are determined by the
average distance $L/N.$ At low densities, the interdisk spacings are large
and uncorrelated, so that again entropy-wise the correlations are connected
to $L/N$ rather than to the global system's size. But at the intermediate
densities, when the longitudinal and nascent transverse orders compete, the
system tends to benefit from both interdisk spaces, the strict $L/N$ and
those nearby $L/N\sim 1.$ To do so, it searches for the space by correlating
interdisk spaces along the system so that the system size comes into play.
As a result, the size effect can manifest itself in the pressure: slight
increase of density adjusts the pressure in an infinite system to that in a
finite one.

\section{Conclusion}

\label{Sec5}

We derived the formulae for the two important PDFs $\,g(R)\,$ and $%
\,g_{1}(R)\,$ for a q1D HD system and demonstrated that they can be readily
used. Apart of that, based on our finding on the correlation lengths, we
suggested that the density $\,\rho =1\,$ plays a distinguished role in the
zigzag transformation with density irrespective of the pore width. We
related this to a nascent longitudinal order and the system tendency to
correlate multiple interdisk spacings along the system to increase its
entropy. To this effect we attributed the high sensitivity of the system
pressure to its density in the vicinity of $\,\rho =1\,$\ which was also
revealed in \cite{Santos1}. As the pressure is affected by a system size and
can be slightly higher in a finite system with periodic boundary conditions
than in an infinite system, the PDF $\,g(R)\,$ and next-neighbor
distribution $\,g_{1}(R)\,$, which nearly coincide with computer simulation
data for high and low densities, can differ for intermediate densities in
the vicinity of $\,\rho =1\,$, but can be made coinciding by the
correspondent density increase in an infinite system. Of course, one obvious
reason for the observed mismatch between the theoretical predictions and
simulation data can be the approximation described in Sec.2, but one cannot
also exclude an effect of the pressure difference between a finite system
with the periodic boundary condition and infinite system, which is possible
at the intermediate densities. Note that the theoretical results \cite%
{Comment,Santos2} based on the transfer martrix approach show a good fit to
the simulation data for high, low, and intermediate densities. As the
periodic boundary conditions along the pore are essential for both these
approaches, the relation between the results obtained for a finite and
infinite systems is yet to be clarified. The investigation of a similar
problem in the physics of one-dimensional ultra cold quantum gases shows
that this problem is nontrivial and worth to be addressed \cite{JofPhysics}. 

\section*{Declaration of competing interest}

The authors declare that they have no known competing financial interests or
personal relationships that could have appeared to influence the work
reported in this paper.

\section*{Acknowledgments}

V.M.P. is grateful to Center for Theoretical Physics PAS for hospitality. V.M.P.'s research is part of the project No. 2022/45/P/ST3/04237 co-funded by the National Science Centre and the
European Union Framework Programme for Research and Innovation Horizon 2020
under the Marie Sk\l odowska-Curie grant agreement No. 945339.
T.B. and A.T. were supported by the National Research Foundation of Ukraine
under the grant 2020.02/0115.






\end{document}